\documentclass[11pt,a4paper]{article}
\usepackage[hyperref]{acl2020}
\usepackage{times}
\usepackage{latexsym}

\newcommand{\mysubsection}[1]{\vspace{0.5em} \noindent\textbf{#1}}
\usepackage{graphicx}
\usepackage{booktabs}

\usepackage[ruled,vlined]{algorithm2e}

\usepackage{microtype}

\aclfinalcopy 

\title{Expressive Interviewing: \\A Conversational System for Coping with COVID-19}

\author{Charles Welch\(^\diamond\), Allison Lahnala\(^\diamond\), Verónica Pérez-Rosas\(^\diamond\), Siqi Shen\(^\diamond\), Sarah Seraj\(^\Phi\),\\ \textbf{Larry An\(^\iota\), Kenneth Resnicow\(^\dagger\), James Pennebaker\(^\Phi\), Rada Mihalcea\(^\diamond\)}\\
\(^\diamond\)Computer Science \& Engineering, University of Michigan\\
\(^\Phi\)Department of Psychology, University of Texas \\
\(^\iota\) Medical School, University of Michigan \\
\(^\dagger\) School of Public Health, University of Michigan\\
\texttt{\{cfwelch,alcllahn,vrncapr,shensq,lcan,kresnic,mihalcea\}@umich.edu} \\ \texttt{\{sarahseraj,pennebaker\}@utexas.edu}}

\date{}

\begin{document}
\maketitle
\begin{abstract}
The ongoing COVID-19 pandemic has raised concerns for many regarding personal and public health implications, financial security and economic stability. Alongside many other unprecedented challenges, there are increasing concerns over social isolation and mental health. We introduce \textit{Expressive Interviewing}--an interview-style conversational system that draws on ideas from motivational interviewing and expressive writing. 
Expressive Interviewing seeks to encourage users to express their thoughts and feelings through writing by asking them questions about how COVID-19 has impacted their lives. 
We present relevant aspects of the system's design and implementation as well as quantitative and qualitative analyses of user interactions with the system. In addition, we conduct a comparative evaluation with a general purpose dialogue system for mental health that shows our system potential in helping users to cope with COVID-19 issues.   
\end{abstract}

\section{Introduction}

The COVID-19 pandemic has changed our world in unimaginable ways, dramatically challenging our health system and drastically changing our daily lives. As we learned from recent large-scale analyses that we performed on social media datasets and extensive surveys, many people are currently experiencing increased anxiety, loneliness, depression, concerns for the health of family and themselves, unexpected unemployment, increased child care or homeschooling, and general concern with what the future might look like.\footnote{\url{http://trackingsocial.life}}


Research in Expressive Writing \cite{Pennebaker97Writing} and Motivational Interviewing \cite{miller2012motivational} has shown that even simple interactions where people talk about one particular experience can have significant psychological value. Numerous studies have demonstrated their effectiveness in improving people’s mental and physical health \cite{Vine20Feelings,pennebaker2011expressive,Resnicow17Efficient}. Both Expressive Writing and Motivational Interviewing rely on the fundamental idea that by putting emotional upheavals into words, one can start to understand them better and therefore gain a sense of agency and coherence of the thoughts and emotions surrounding their experience. 


In this paper, we introduce a new interview-style dialogue paradigm called \textit{Expressive Interviewing} that unites strategies from Expressive Writing and Motivational Interviewing through a system that guides an individual to  reflect on, express, and better understand their own thoughts and feelings during the pandemic. 

By encouraging introspection and self-expression, the dialogue aims to reduce stress and anxiety. 
Our system is currently online at \url{https://expressiveinterviewing.org} and available for anyone to try anonymously.

\section{Related Work}

\mysubsection{Expressive Writing.} 
Expressive writing is a writing paradigm where people are asked to disclose their emotions and thoughts about significant life upheavals. Originally studied in the scope of traumatic experiences~\cite{pennebaker1986confronting}, study participants are usually asked to write about an assigned topic for about 15 minutes for one to five consecutive days. Later studies expanded to specific experiences such as losing a job~\cite{spera1994expressive}.
Expressive writing has been shown to be effective on both physical and mental health measures by multiple meta-analyses~\cite{frattaroli2006experimental}, finding its association with drops in physician visits, positive behavioral changes, and long-term mood improvements.
No single theory at present explains the cause of its benefits, but it is believed that the process of expressing emotions and constructing a story may play a role for participants in forming a new perspective on their lives~\cite{pennebaker2011expressive}. 

\mysubsection{Motivational Interviewing.}
Motivational Interviewing (MI) is a counseling technique designed to help people change a desired behavior by leveraging their own values and interests. The approach accepts that many people looking for a change are ambivalent about doing so as they have reasons to both change and sustain the behavior. Therefore, the goal of an MI counselor is to elicit their client's own motivation for changing by asking open questions and reflecting back on the client's statements.
MI has been shown to correlate with positive behavior changes in a large variety of client goals, such as weight management \cite{small2009pediatric}, chronic care intervention \cite{brodie2008motivational}, and substance abuse prevention \cite{d2008brief}.

\mysubsection{Dialogue Systems.}
With the development of deep learning techniques, dialogue systems have been applied to a large variety of tasks to meet increasing demands. In recent work, \citet{afzal-etal-2019-development} built a dialogue-based tutoring system to guide learners through varying levels of content granularity to facilitate a better understanding of content. \citet{henderson-etal-2019-polyresponse} applied a response retrieval approach in restaurant search and booking to provide and enable the users to ask various questions about a restaurant. \citet{ortega-etal-2019-adviser} built an open-source dialogue system framework that navigates students through course selection. 

There are also dialogue system building tools such as Google's Dialogflow\footnote{\url{https://dialogflow.com/}} and IBM's Watson assistant,\footnote{\url{https://www.ibm.com/cloud/watson-assistant}} which enable numerous dialogue systems for customer service or conversational user interfaces.

\mysubsection{Chatbots for Automated Counseling.}
Two dialogue systems for automated counseling services available on mobile platforms are Wysa\footnote{\url{https://wysa.io/}} and Woebot.\footnote{\url{https://woebot.io/}}
These chatbots provide cognitive behavioral therapy with the goal of easing anxiety and depression by allowing users to express their thoughts. A study of Wysa users over three months showed that more active users had significantly improved symptoms of depression~\cite{inkster2018empathy}. Another study shows that young students using Woebot significantly reduced anxiety levels after two weeks of using the conversational agent \cite{fitzpatrick2017delivering}. These findings suggest a promising benefit of automated counseling for the nonclinical population.


Our system is distinct from Wysa and Woebot in that it is designed specifically for coping with COVID-19 and allows users to write more topic related free-form responses. It asks open-ended questions and encourages users to introspect, and then provides visualized feedback afterward, whereas the others have a conversational logic mainly based on precoded multiple choice options.

\section{Expressive Interviewing}

Our system conducts an interview-style interaction with the users about how the COVID-19 pandemic has been affecting them. The interview consists of several writing prompts in the form of questions about specific issues related to the pandemic. During the interview, the system provides reflective feedback based on the user's answers. After the interaction is concluded, the system presents users with detailed graphical and textual feedback. 

The system's goal is to encourage users to 
write as much as possible about themselves, building upon previous findings regarding the psychological value of writing about personal upheavals and the use of reflective listening  for behavioral change \cite{Pennebaker97Writing,miller2012motivational}. To achieve this, the system guides the interaction by 
asking four main open-ended questions. Then, based on users responses, the system provides feedback and asks additional questions whenever appropriate. In order to provide reflective feedback, the system automatically detects the topics being discussed (e.g., work, family) or emotions being felt (e.g., anger, anxiety), and responds with a reflective prompt that asks the user to elaborate or to answer a related question to explore that concept more deeply. For instance, if the system detects \emph{work} as a topic of interest, it responds with ``How has work changed under COVID? What might you be able to do to keep your career moving during these difficult times?''

\subsection{Leading Questions}

During the formulation of the guiding questions used by our system, we worked closely with our psychology and public health collaborators to identify a set of questions on COVID-19 topics that would motivate individuals to talk about their personal experience with the pandemic. 
We formulated the following question as the system's conversation starting point:
\vskip 0.1in
\noindent {\bf [Major issues]} What are the major issues in your life right now, especially in the light of the COVID outbreak?
\vskip 0.1in
We also formulated three follow-up questions, which were generated after several refining iterations.\footnote{We removed an additional question about how people's lives have changed since the outbreak, as well as a question about what people missed the most about their previous lives.} The order of these questions is randomized across users of the system.

\vskip 0.1in
\noindent {\bf [Looking Forward]} What do you most look forward to doing once the pandemic is over?
\vskip 0.1in
\noindent {\bf [Advice to Others]} What advice would you give other people about how to cope with any of the issues you are facing?
\vskip 0.1in
\noindent {\bf [Grateful]} The outbreak has been affecting everyone's life, but people have the amazing ability to find good things even in the most challenging situations. What is something that you have done or experienced recently that you are grateful for?\\

\subsection{Language Understanding and Reflection Strategies}

Our system's capability for language understanding relies on identifying words belonging to various lexicons. This simple strategy allowed us to quickly develop a platform upon which we intend to implement a more sophisticated language understanding ability in future work.

When a user responds to one of the main prompts, the system looks for words belonging to specific topics and word categories. The system examines the user responses to identify dominant word categories or topics and  
triggers a reflection from a set of appropriate reflections.~\footnote{A dominant word category is defined as a word type, where the frequency of occurrence is at least 50\% higher than the second highest frequency category for that group.} If none of these types are matched, it responds with a generic reflection.

The word categories are derived from the LIWC, WordNet-Affect and MPQA lexicons~\cite{pennebaker2001linguistic,strapparava2004wordnet,wiebe2005annotating} and include pronouns (I, we, others), negative emotion (anger, anxiety, and sadness), positive emotion (joy) and positive and negative words. The COVID-19 related topics include finances, health, home, work, family, friends, and politics. Most of the topics are covered by the LIWC lexicon, with the exception of politics. For this category, we use the \emph{politics} category from the Roget's Thesaurus~\cite{roget1911roget} and add a small number of proper nouns covered in recent news (e.g. Trump, Biden, Fauci, Sanders).

We formulate a set of specific reflections for each word category and topic, which were refined by our psychology and public health collaborators. For instance, if the dominant emotion category is anxiety, the system responds ``You mention feelings such as fear and anxiety. What do you think is the best way for people to cope with these feelings?''
Initially, we also considered reflections for different types of pronouns, but found that they did not steer the dialogue in a meaningful direction. Instead, we flag responses with dominant use of impersonal pronouns and lack of references to the self and reflect that fact back to the user and further ask them
how they are specifically being affected. 
We also crafted generic reflections to be applicable to a large number of situations though the system does not understand the content of what the user has said (e.g. ``I see. Tell me about a time when things were different'', and ``I hear you. What have you tried in the past that has worked well'').

\subsection{User Feedback}

After the interview, the system
provides visual and textual feedback based on the user's responses and provides links to  resources (i.e., mental health resources) appropriate given their main concerns.

The visual feedback consists of four pie charts showing the relative usage of different word categories, including: discussed topics (work, finance, home, health, family, friends and politics), 
affect (positive, negative), emotions (anger, sadness, fear, anxiety, joy), and pronouns (I, we, other).


The textual feedback includes a comparison with others (to normalize the user's reactions) and interpretations of where the user falls within normalized scales. The system also presents a summary of the most and least discussed topics and how they compare to the average user, along with 
normalized values for meaningfulness, self-reflection, and emotional tone (using a 0-10 scale) along with textual descriptors for the shown scale values.~\footnote{Textual descriptions are predefined for different ranges of each scale}  
These metrics are inspired by previous work on expressive writing and represent the self-reported meaningfulness, usage of self-referring pronouns, and the difference in positive and negative word usage~\cite{pennebaker1997writing}.
Finally, the system provides relevant resources for further exploration (e.g. for the work topic it lists external links to COVID related job resources and safety practices).

\subsection{Online Interface}
The system is implemented as a web interface so it is accessible and easy to use.  The 
interface is built with the Django platform and jQuery and uses Python on the backend~\cite{django}. 

{\bf Before} the interaction 
users are asked to report on a 1-7 scale: (1) {\bf [Life satisfaction]} how satisfied they are with their life in general, and (2) {\bf [Stress$_{before}$]} what is their level of stress. The user then proceeds to the conversational interaction with our system. {\bf After} the interaction, the user is asked again about (3) {\bf [Stress$_{after}$]} what is their level of stress;  (4) {\bf [Personal]} how personal their interaction was; and (5) {\bf [Meaningful]} how meaningful their interaction was. Once this is submitted, the user can proceed to the feedback page and view details about what they wrote and how their interaction compares to a sample of recent users. The user is finally presented with a list of resources triggered by the topics discussed.


We made an effort to make our system appear human-like to make users more comfortable while interacting with it, although this can vary for different individuals. In future work, we  hope to explore individual personas and more sophisticated rapport building techniques. We named our dialogue agent `C.P.', which stands for \emph{Computer Program}. This name acknowledges that the user is interacting with a computer, while at the same time it makes the system more human by assigning it a name. When responding to the user, C.P. pauses for a few seconds as if it is thinking and then proceeds to type a response one letter at a time with a low probability of making typos -- similarly to how human users would type.


\section{Analysis of User Interactions}

After the system was launched (and up to when we conducted this analysis), we had 174 users interact with the system. We analyze these interactions to evaluate system usefulness, user engagement, and reflection effectiveness. 

\mysubsection{System Usefulness.} We examine the system's ability to help users cope with COVID-19 related issues by analyzing the different ratings provided by users before and after their interaction with C.P. Throughout this discussion, we use $\Delta$Stress to indicate how the users stress rating differs before and after the interaction: $\Delta$Stress = Stress$_{after}$ - Stress$_{before}$. Negative values for $\Delta$Stress are therefore an indicator of stress reduction, whereas positive values for $\Delta$Stress reflect an increase in stress. 

We start by measuring the Spearman correlation between the different ratings for the 174 interactions with C.P. Results are shown in Table~\ref{tab:rating-correlations}. 


The strongest correlation we observe is between the \textit{personal} and \textit{meaningful} ratings, suggesting that interactions that are more \textit{meaningful} appear to feel more \textit{personal}, or vice versa.

We also observe a strong negative correlation between  $\Delta$ Stress and the meaningfulness of the interaction, suggesting that the interactions that the users found to be meaningful are associated with a reduction in stress. 

\begin{table}[t]
    \centering
    \begin{tabular}{llr}
    \toprule
    Rating 1 & Rating 2 & rho \\
    \midrule
    Life satisfaction & Stress$_{before}$ & \textbf{-0.261} \\
    Life satisfaction & Stress$_{after}$ & \textbf{-0.166} \\ 
    Life satisfaction & $\Delta$ Stress & 0.083 \\
    Life satisfaction & Personal & \textbf{0.285} \\
    Life satisfaction & Meaningful & \textbf{0.243} \\
    \midrule
   Meaningful & Stress$_{before}$ &  0.033 \\
    Meaningful & Stress$_{after}$  & \textbf{-0.226} \\
    Meaningful & $\Delta$ Stress & \textbf{-0.202} \\
     Meaningful & Personal &  \textbf{0.675} \\
    \midrule
    Personal & Stress$_{before}$ &  0.065 \\
    Personal & Stress$_{after}$ & -0.067 \\
    Personal & $\Delta$ Stress & -0.073 \\
    \bottomrule
    \end{tabular}
    \caption{Spearman correlation coefficients between pairs of ratinga for the 174 interactions. Bold indicates significance with $p<0.05$.}
    \label{tab:rating-correlations}
\end{table}

\begin{figure}[h]
    \centering
    \includegraphics[width=\linewidth]{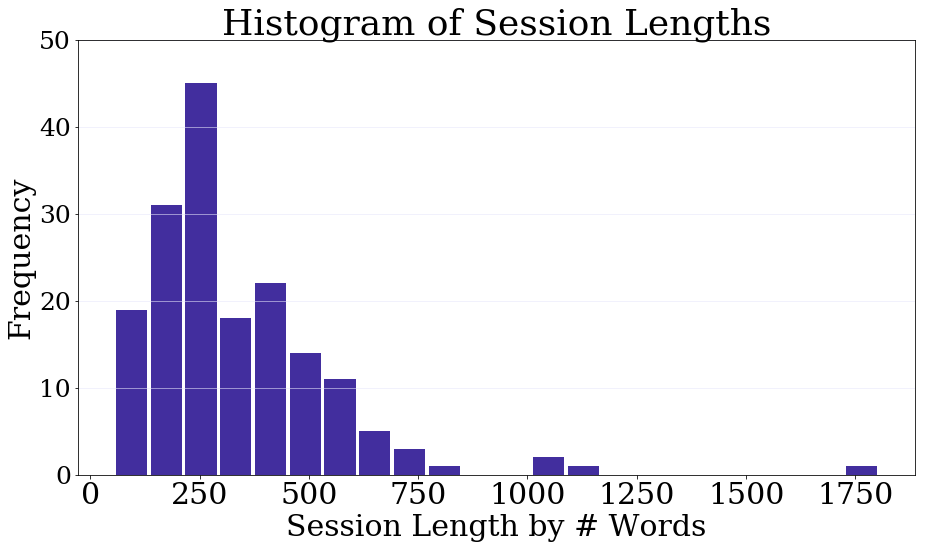}
    \includegraphics[width=\linewidth]{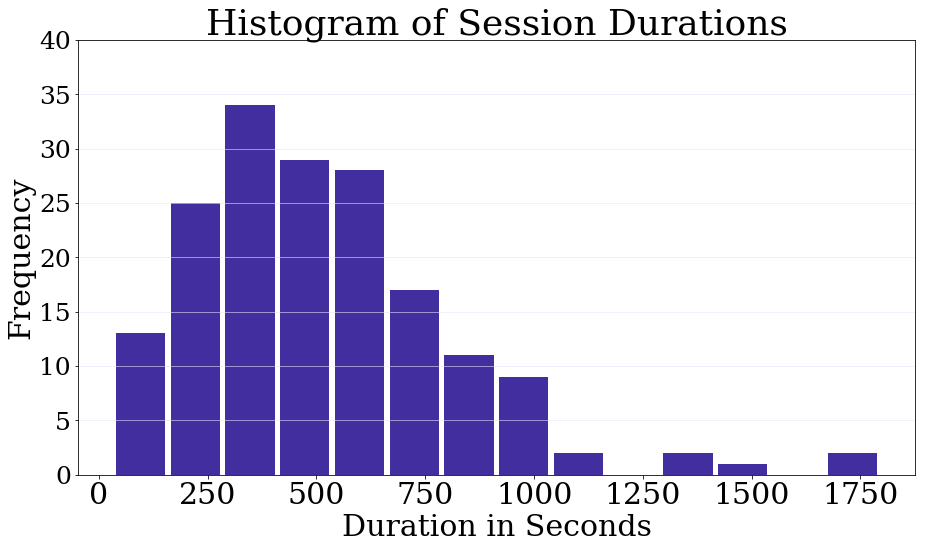}
    \caption{Histograms of overall user engagement measured by session length and duration.}
    \label{fig:length-duration-histogram}
\end{figure}

\mysubsection{User engagement.} We examine user engagement by analyzing the time users spend in the interaction and the number of words they write throughout the session. Figure~\ref{fig:length-duration-histogram} shows histograms of the session lengths in the number of words used by the user and of the session duration in seconds.
The rightmost column of Table~\ref{tab:prompt-rating-correlations} shows Spearman correlation coefficients between user ratings and the length and duration of the sessions. We find a significant negative correlation between  \textit{Stress$_{before}$} and \textit{Stress$_{after}$} with session duration and number of words, suggesting an association between user engagement and lower stress. There is also a weak negative correlation between duration of session and reduction in stress ($\Delta$Stress). 


\begin{table*}[h]
\centering
\begin{tabular}{llllll}
 \hline
 & \multicolumn{1}{c}{Major issues} & \multicolumn{1}{c}{Grateful} & \multicolumn{1}{c}{Looking Forward} & \multicolumn{1}{c}{Advice to Others} & Overall \\
 
                 
                \cmidrule{2-6}
                 & \multicolumn{4}{c}{Length in Words}                                                                                                                   \\ \midrule
Life Satisfaction   & -0.148 & -0.121 & -0.079 & -0.096 & -0.070 \\
Personal    & \textbf{-0.156} & \textbf{-0.147} & \textbf{-0.185} & -0.134 & \textbf{-0.159} \\
Meaningful  & \textbf{-0.181} & \textbf{-0.148} & -0.142 & \textbf{-0.151} & \textbf{-0.151} \\
Stress$_{before}$    & -0.001 & -0.076 & \textbf{-0.161} & -0.083 & \textbf{-0.151} \\
Stress$_{after}$     & -0.020 & -0.135 & -0.130 & -0.129 & \textbf{-0.177} \\
$\Delta$ Stress & -0.067 & -0.106 & -0.039 & -0.112 & -0.092 \\
                \midrule
                 & \multicolumn{4}{c}{Duration in Seconds} \\\midrule
Life Satisfaction        & -0.057 & 0.016 & 0.048 & 0.091 & 0.066 \\
Personal         & -0.017 & -0.041 & 0.053 & 0.136 & 0.066 \\
Meaningful       & -0.036 & 0.099 & \textbf{0.173} & \textbf{0.205} & 0.143 \\
Stress$_{before}$      & -0.067 & \textbf{-0.252} & \textbf{-0.178} & -0.099 & \textbf{-0.198} \\
Stress$_{after}$      & -0.120 & \textbf{-0.241} & \textbf{-0.207} & \textbf{-0.192} & \textbf{-0.233} \\
$\Delta$ Stress & -0.069 & -0.023 & -0.052 & -0.092 & -0.068 \\
\bottomrule                                 
\end{tabular}
\caption{Spearman correlation coefficients between each rating provided by a user and (top) the length in number of words of the user's response to each particular prompt, and (bottom) duration in seconds of the user's response to each particular prompt, from 174 full interactions. Bold denotes significance with $p<0.05$.}
    \label{tab:prompt-rating-correlations}
\end{table*}

We also investigate if there is a relationship between the pre- and post-session ratings and how engaged a user was with each prompt in terms of length of and duration in writing their response. Table~\ref{tab:prompt-rating-correlations} shows Spearman correlation coefficients for these relationships. 
It appears that \textit{Life Satisfaction} has no correlation with the length of any prompt response except a potentially weak negative correlation with length on the \textit{Major Issues} prompt ($p=0.052$). A lower rating may relate with having more personal challenges to write about.

\textit{Stress$_{before}$} has a weak negative correlation between the number of words used and the duration spent in the response to \textit{Looking Forward}. Higher stress may relate to present concerns,  which may make one less inclined to spend time thinking and writing about positive aspects of their future than someone with less stress.
We presume this could be the case for the \textit{Grateful} prompt, which likewise correlates weakly and negatively with \textit{Stress$_{before}$}.

\textit{Stress$_{after}$} has a negative correlation between duration spent on every prompt response except for the time spent on \textit{Major Issues}. This could be a reflection of the fact that those who have a lot to write about major issues in their life also incur high levels of stress. 

The \textit{Personal} rating shows no correlations with the duration spent on any of responses, except potentially \textit{Advice to Others} ($p=0.074$). We do observe weak negative correlations between \textit{Personal} ratings and response \textit{lengths} on \textit{Major Issues} and \textit{Looking Forward}, and potentially on \textit{Grateful} ($p=0.054$) and \textit{Advice to Others} ($p=0.08$). Perhaps if a user writes more, there is a greater expectation for more personal reflections. We discuss engagement related to reflections more deeply in the next section.

The \textit{Meaningful} rating shows weak negative correlations with length on \textit{Major Issues}, \textit{Advice to Others}, and possibly on \textit{Grateful} ($p=0.052$) and \textit{Looking Forward} ($p=0.062$). We do not observe a significant correlation with \textit{duration} on \textit{Major Issues} or \textit{Grateful}, but we do observe positive correlations between \textit{duration} and \textit{Looking Forward} and \textit{Advice to Others}. Users who spend more time thinking about advice they would give others facing their issues may find the interaction more meaningful, and may experience benefits having reflected on their agency in managing their challenges.

\begin{table*}[t]
    \centering
    \begin{tabular}{p{1.5cm}p{12.5cm}}
\hline 
HEALTH & I’d like to know more about your feelings surrounding your own health and the health of people close to you. What actions can you take to help keep you healthy during these challenging times? \\
FAMILY &  What can you do to keep your family resilient during these tough times? \\
POLITICS & What is it about the political world that may be hooking you? What are your reactions saying about you? \\
GEN1 & Interesting to hear that. How does what you say relate to your values? \\
GEN3 & I see. Tell me about a time when things were different. \\
\hline 
    \end{tabular}
    \caption{Sample topic specific and generic reflections.}
    \label{tab:sample_reflections}
\end{table*}

\begin{figure}[h]
    \centering
    \includegraphics[width=\linewidth]{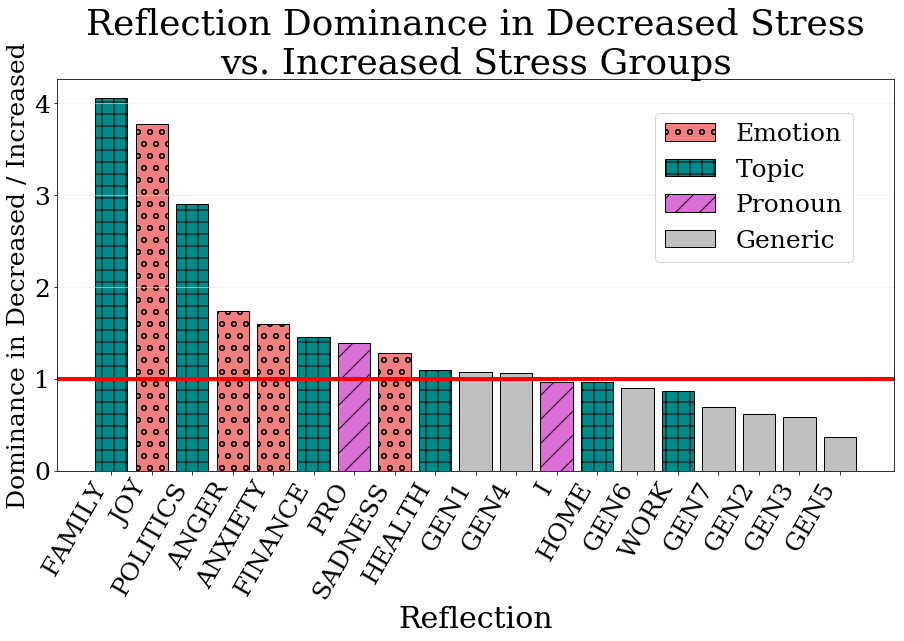}
    \caption{The dominance of each reflection triggered for users whose stress decreased divided by each reflection's dominance for users whose stress increased. Scores above 1 (red line) correspond to a decrease in stress; score below 1 correspond to an increase in stress. See Table \ref{tab:sample_reflections} for sample reflections, including the {\sc gen}eric reflections.}
    \label{fig:reflection-dominance}
\end{figure}

\mysubsection{Reflection Effectiveness.} To investigate the effectiveness of Expressive Interviewing reflections, we compare the reflections that were triggered for users whose stressed decreased to the reflections that triggered for the users whose stress increased. For each of these user groups, we compute the dominance of each reflection as its proportion of times it was triggered out of all reflections triggered. In Figure~\ref{fig:reflection-dominance}, we compare the dominance of each reflection across these user groups by dividing the reflection dominance in the decreased-stress group by that of the increased-stress group.

Importantly, we observe that all emotion reflections and more topic reflections were triggered at a higher rate for users whose stress decreased, whereas more generic reflections were triggered at a higher rate for users whose stress increased. While we do not presume that increased stress was due to generic reflections, the correspondence between emotion and topic reflections with stress reduction aligns with expectations of effective reflections from Motivational Interviewing--generic reflections and specific reflections  resemble \textit{simple reflections} and \textit{complex reflections} respectively, as referred to  in Motivation Interviewing. While both types of reflections serve a purpose, complex reflections both communicate an understanding of what the client has said and also contribute an additional layer of understanding or a new interpretation for the user, whereas simple reflections focus on the former~\cite{rollnick2004motivational}.

 In qualitatively analyzing the instances where generic reflections were triggered, we observe that contextual appropriateness seems to be the best indicator of their success (in terms of ability to elicit a deeper thought, feeling, or interpretation) given that the user was invested in the experience. 
 As these generic reflections are selected at random, their contextual appropriateness was inconsistent, illuminating the scenarios in which they are more or less appropriate.
 For instance, out of the seven times the reflection ``\textit{Interesting to hear that. How does what you say relate to your values?}" was triggered for the increased-stress users, one user expanded on their previous message, one expressed confusion about the question, and another copied and pasted the definition of \textit{core values}\footnote{\url{https://www.indeed.com/career-advice/career-development/core-values}} as their response. Two other instances of this reflection were triggered when a user had expressed negative feelings such as worry and feeling lazy which appeared misplaced, and the last case was triggered by a message that was not readable. Out of the thirteen times the same reflection was triggered for the decreased-stress group, one user expressed not having much to say, another gave one word responses before and after, and all others expanded on their previous message in relation to their values or gave a simple response to indicate a degree that it relates. This reflection appeared more ``successful" (based on if the user expanded on their previous message or values) when it was triggered by a message with more neutral to positive sentiment, such as when the user was expressing what they were looking forward to, or when they had several pieces of advice to offer for a friend in their situation, as opposed to one with more negative sentiment like the messages expressing worry or laziness. 
 
 In instances of other generic reflections, we observed that another issue for appropriateness was whether the reflection matched the user's frame of thought in terms of past, present, or future. For instance, the reflection \textit{``I see. Tell me about a time when things were different,"} best matched scenarios when users described thoughts about changes to their daily lives, but not when users described future topics such as what they were looking forward to, nor when they were already describing the past.
 
 
Based on our observations of the reflections in action, we have three main takeaways. First, topic and emotion specific reflections are more associated with the group of users whose stress decreased. These reflections are only triggered if the system determines a dominant topic or emotion, which depends on the effectiveness of its heuristics, as well as the amount of detail and context that a user expresses. This leads to the next takeaway, that the system appears to be more effective when users approach the experience with an intention for expression, or conversely it seems less effective when the intent to not engage and express is explicit. Third, the generic reflections were developed with the intent to function in generic contexts, but we learned in practice that some clashed with emotional and situational content or were confusing given the context. As we did observe many, if not more, successful instances of generic reflections, we are able to contrast these contexts to the unsuccessful contexts, and can develop a heuristic for selecting the generic reflections rather than selecting at random, as well as adapt the language of our current generic reflections to be more appropriate for the Expressive Interviewing setting.




\section{Comparative Evaluations}
To assess the extent to which our Expressive Interviewing system delivers an engaging user experience, we conduct a comparative study between our system and the conversational mental health app Woebot~\cite{fitzpatrick2017delivering}. 

We recruited 12 participants and asked them to interact independently with each system to discuss their COVID-19 related concerns. More specifically, we asked them to use each system for 10-15 minutes and provide evaluative feedback pre- and post-interaction. To avoid cognitive bias, we randomized the order in which each participant evaluated the systems. In addition, we randomized the order in which the evaluation questions are shown. 

Before interacting with either system, participants rated their life satisfaction and their stress level. After the interaction, participants reported again their stress level and rated several aspects of their interaction with the system, including ease of use, usefulness (in terms of discussing COVID-19 related issues and motivation to write about it), overall experience, and satisfaction using mainly binary scales. For example, the questions ``Did $<$system$>$ motivate you to write at length about your thoughts and feelings? yes/no'' and 
``How useful was C.P. to discuss your concerns about COVID? useful/not useful" assess whether the system encouraged the user to write about their thoughts and feelings about COVID and whether the system provided guidance for it. Tables~\ref{tab:evaluation-stress} and~\ref{tab:evaluation} show the percentage of users that provided positive or high scores ($>3$ on a 7-point scale) for each of these aspects after interacting with both systems. 

\begin{table}[t]
    \centering
    \begin{tabular}{lcc}
    \toprule
    	&	Woebot	&	Expressive Interviewing	\\ \midrule
Stress$_{before}$	&	91\%	&	91\%	\\
Stress$_{after}$	&	73\%	&	64\%	\\ \bottomrule
\end{tabular}
    \caption{Percentage of users reporting high levels of stress ($>3$ on a 7-point Likert scale) before and after using Woebot and Expressive Interviewing.}
    \label{tab:evaluation-stress}
\end{table}

As observed, there are fewer participants reporting high levels of stress after using either system. However, we see a smaller fraction of participants reporting high levels of stress after interacting with Expressive Interviewing, thus suggesting that our system was more effective in helping participants to reduce their stress levels. 

Overall, participants reported that Expressive Interviewing was easier to use, more useful to discuss their COVID concerns and motivated them to write more than Woebot. Similarly, users reported a more meaningful interaction and a better overall experience. However, it is important to mention that Woebot was not specifically designed for discussing COVID-19 concerns and it is of more general purpose than our system. Nonetheless, we believe that this comparison provides evidence that a dialogue system such as Expressive Interviewing is more effective in helping users cope with COVID-19 issues as compared to a general purpose dialogue system for mental health.

\begin{table}[t]
    \centering
    \begin{tabular}{lcc}
\toprule
	&	Woebot	&	Expr. Interv.	\\ \midrule
Ease of Use	&	82\%	&	91\%	\\
Useful	&	18\%	&	73\%	\\
Motivation to Write	&	27\%	&	91\%	\\
User Satisfaction	&	36\%	&	36\%	\\
Meaningful Interaction 	&	64\%	&	73\%	\\
Overall Experience 	&	36\%	&	46\%	\\
\bottomrule
\end{tabular}
    \caption{Comparative evaluation Woebot and Expressive Interviewing. Percentage of users reporting positive/high ratings (with scores $>$3 in a 7-point Likert scale) on usability aspects after interacting with Woebot and Expressive Interviewing.}
    \label{tab:evaluation}
\end{table}





\section{Ethical and Privacy Considerations}

We followed the suggestions of previous research on automated mental health counseling and adopted the goals of being respectful of user privacy, following evidence based methods, ensuring user safety, and being transparent in system capabilities~\cite{kretzschmar2019can}. The practices of motivational interviewing and expressive writing have numerous studies supporting their efficacy~\cite{miller2012motivational,pennebaker2007expressive}. The combination of these methods in an interviewing format has not previously been studied and we intend to continue publishing our findings as the user population expands and becomes more diverse. We will also continue to improve our system and  assessment.

We have taken efforts to secure user data. We do not ask for identifiers and data is stored anonymously by session ID. 
The website is secured with SSL. Data is only accessible to researchers directly involved with our study.

Our study has been approved by the University of Michigan IRB. 


\section{Conclusion}
In this paper, we introduced an interview-style dialogue system called Expressive Interviewing to help people cope with the effects of the COVID-19 pandemic. We provided a detailed description on how the system is designed and implemented. 

We analyzed a sample of 174 user interactions with our system and conducted qualitative and quantitative analyses on aspects such as system usefulness, user engagement and reflection effectiveness. We also conducted a comparative evaluation study between our system and Woebot, a general purpose dialogue system for mental health. Our main findings suggest that users benefited from the reflective strategies used by our system and experienced meaningful interactions leading to reduced stress levels. Furthermore, our system was judged to be easier to use and more useful than Woebot when discussing COVID-19 related concerns.  

In future work we intend to explore the applicability of the developed system to other health-related domains. 

\section*{Acknowledgements}
This material is based in part upon work supported by the Precision Health initiative at the University of Michigan, by the National Science Foundation (grant \#1815291), and by the John Templeton Foundation (grant \#61156). Any opinions, findings, and conclusions or recommendations expressed in this material are those of the authors and do not necessarily reflect the views of the Precision Health initiative, the National Science Foundation, or John Templeton Foundation.



\bibliography{main}

\begin{thebibliography}{25}
\expandafter\ifx\csname natexlab\endcsname\relax\def\natexlab#1{#1}\fi

\bibitem[{Afzal et~al.(2019)Afzal, Dhamecha, Mukhi, Sindhgatta, Marvaniya,
  Ventura, and Yarbro}]{afzal-etal-2019-development}
Shazia Afzal, Tejas Dhamecha, Nirmal Mukhi, Renuka Sindhgatta, Smit Marvaniya,
  Matthew Ventura, and Jessica Yarbro. 2019.
\newblock \href {https://doi.org/10.18653/v1/N19-2015} {Development and
  deployment of a large-scale dialog-based intelligent tutoring system}.
\newblock In \emph{Proceedings of the 2019 Conference of the North {A}merican
  Chapter of the Association for Computational Linguistics: Human Language
  Technologies, Volume 2 (Industry Papers)}, pages 114--121, Minneapolis,
  Minnesota. Association for Computational Linguistics.

\bibitem[{Brodie et~al.(2008)Brodie, Inoue, and Shaw}]{brodie2008motivational}
David~A Brodie, Allison Inoue, and David~G Shaw. 2008.
\newblock Motivational interviewing to change quality of life for people with
  chronic heart failure: a randomised controlled trial.
\newblock \emph{International journal of nursing studies}, 45(4):489--500.

\bibitem[{D'Amico et~al.(2008)D'Amico, Miles, Stern, and Meredith}]{d2008brief}
Elizabeth~J D'Amico, Jeremy~NV Miles, Stefanie~A Stern, and Lisa~S Meredith.
  2008.
\newblock Brief motivational interviewing for teens at risk of substance use
  consequences: A randomized pilot study in a primary care clinic.
\newblock \emph{Journal of substance abuse treatment}, 35(1):53--61.

\bibitem[{{Django Software Foundation}(2019)}]{django}
{Django Software Foundation}. 2019.
\newblock \href {https://djangoproject.com} {Django}.

\bibitem[{Fitzpatrick et~al.(2017)Fitzpatrick, Darcy, and
  Vierhile}]{fitzpatrick2017delivering}
Kathleen~Kara Fitzpatrick, Alison Darcy, and Molly Vierhile. 2017.
\newblock Delivering cognitive behavior therapy to young adults with symptoms
  of depression and anxiety using a fully automated conversational agent
  (woebot): a randomized controlled trial.
\newblock \emph{JMIR mental health}, 4(2):e19.

\bibitem[{Frattaroli(2006)}]{frattaroli2006experimental}
Joanne Frattaroli. 2006.
\newblock Experimental disclosure and its moderators: a meta-analysis.
\newblock \emph{Psychological bulletin}, 132(6):823.

\bibitem[{Henderson et~al.(2019)Henderson, Vuli{\'c}, Casanueva, Budzianowski,
  Gerz, Coope, Spithourakis, Wen, Mrk{\v{s}}i{\'c}, and
  Su}]{henderson-etal-2019-polyresponse}
Matthew Henderson, Ivan Vuli{\'c}, I{\~n}igo Casanueva, Pawe{\l} Budzianowski,
  Daniela Gerz, Sam Coope, Georgios Spithourakis, Tsung-Hsien Wen, Nikola
  Mrk{\v{s}}i{\'c}, and Pei-Hao Su. 2019.
\newblock \href {https://doi.org/10.18653/v1/D19-3031} {{P}oly{R}esponse: A
  rank-based approach to task-oriented dialogue with application in restaurant
  search and booking}.
\newblock In \emph{Proceedings of the 2019 Conference on Empirical Methods in
  Natural Language Processing and the 9th International Joint Conference on
  Natural Language Processing (EMNLP-IJCNLP): System Demonstrations}, pages
  181--186, Hong Kong, China. Association for Computational Linguistics.

\bibitem[{Inkster et~al.(2018)Inkster, Sarda, and
  Subramanian}]{inkster2018empathy}
Becky Inkster, Shubhankar Sarda, and Vinod Subramanian. 2018.
\newblock An empathy-driven, conversational artificial intelligence agent
  (wysa) for digital mental well-being: real-world data evaluation
  mixed-methods study.
\newblock \emph{JMIR mHealth and uHealth}, 6(11):e12106.

\bibitem[{Kretzschmar et~al.(2019)Kretzschmar, Tyroll, Pavarini, Manzini,
  Singh, and Group}]{kretzschmar2019can}
Kira Kretzschmar, Holly Tyroll, Gabriela Pavarini, Arianna Manzini, Ilina
  Singh, and NeurOx Young People’s~Advisory Group. 2019.
\newblock Can your phone be your therapist? young people’s ethical
  perspectives on the use of fully automated conversational agents (chatbots)
  in mental health support.
\newblock \emph{Biomedical informatics insights}, 11:1178222619829083.

\bibitem[{Miller and Rollnick(2012)}]{miller2012motivational}
William~R Miller and Stephen Rollnick. 2012.
\newblock \emph{Motivational interviewing: Helping people change}.
\newblock Guilford press.

\bibitem[{Ortega et~al.(2019)Ortega, V{\"a}th, Weber, Vanderlyn, Schmidt,
  V{\"o}lkel, Karacevic, and Vu}]{ortega-etal-2019-adviser}
Daniel Ortega, Dirk V{\"a}th, Gianna Weber, Lindsey Vanderlyn, Maximilian
  Schmidt, Moritz V{\"o}lkel, Zorica Karacevic, and Ngoc~Thang Vu. 2019.
\newblock \href {https://doi.org/10.18653/v1/P19-3016} {{ADVISER}: A dialog
  system framework for education {\&} research}.
\newblock In \emph{Proceedings of the 57th Annual Meeting of the Association
  for Computational Linguistics: System Demonstrations}, pages 93--98,
  Florence, Italy. Association for Computational Linguistics.

\bibitem[{Pennebaker(1997{\natexlab{a}})}]{pennebaker1997writing}
James~W Pennebaker. 1997{\natexlab{a}}.
\newblock Writing about emotional experiences as a therapeutic process.
\newblock \emph{Psychological science}, 8(3):162--166.

\bibitem[{Pennebaker and Beall(1986)}]{pennebaker1986confronting}
James~W Pennebaker and Sandra~K Beall. 1986.
\newblock Confronting a traumatic event: toward an understanding of inhibition
  and disease.
\newblock \emph{Journal of abnormal psychology}, 95(3):274.

\bibitem[{Pennebaker and Chung(2007)}]{pennebaker2007expressive}
James~W Pennebaker and Cindy~K Chung. 2007.
\newblock Expressive writing, emotional upheavals, and health.
\newblock \emph{Foundations of health psychology}, pages 263--284.

\bibitem[{Pennebaker and Chung(2011)}]{pennebaker2011expressive}
James~W Pennebaker and Cindy~K Chung. 2011.
\newblock Expressive writing: Connections to physical and mental health.
\newblock Oxford University Press.

\bibitem[{Pennebaker et~al.(2001)Pennebaker, Francis, and
  Booth}]{pennebaker2001linguistic}
James~W Pennebaker, Martha~E Francis, and Roger~J Booth. 2001.
\newblock Linguistic inquiry and word count: Liwc 2001.
\newblock \emph{Mahway: Lawrence Erlbaum Associates}, 71(2001):2001.

\bibitem[{Pennebaker(1997{\natexlab{b}})}]{Pennebaker97Writing}
J.W. Pennebaker. 1997{\natexlab{b}}.
\newblock Writing about emotional experiences as a therapeutic process.
\newblock \emph{Psychological Science}, (8):162--166.

\bibitem[{Resnicow et~al.(2017)Resnicow, Teixeira, and
  Williams}]{Resnicow17Efficient}
K.~Resnicow, PJ~Teixeira, and GC~Williams. 2017.
\newblock Efficient allocation of public health and behavior change resources:
  The "difficulty by motivation" matrix.
\newblock \emph{American Journal of Public Health}, 107(1):55--57.

\bibitem[{Roget(1911)}]{roget1911roget}
Peter~Mark Roget. 1911.
\newblock \emph{Roget's Thesaurus of English Words and Phrases...}
\newblock TY Crowell Company.

\bibitem[{Rollnick and Allison(2004)}]{rollnick2004motivational}
Stephen Rollnick and Jeff Allison. 2004.
\newblock Motivational interviewing.
\newblock \emph{The essential handbook of treatment and prevention of alcohol
  problems}, pages 105--116.

\bibitem[{Small et~al.(2009)Small, Anderson, Sidora-Arcoleo, and
  Gance-Cleveland}]{small2009pediatric}
Leigh Small, Deborah Anderson, Kimberly Sidora-Arcoleo, and Bonnie
  Gance-Cleveland. 2009.
\newblock Pediatric nurse practitioners' assessment and management of childhood
  overweight/obesity: Results from 1999 and 2005 cohort surveys.
\newblock \emph{Journal of Pediatric Health Care}, 23(4):231--241.

\bibitem[{Spera et~al.(1994)Spera, Buhrfeind, and
  Pennebaker}]{spera1994expressive}
Stefanie~P Spera, Eric~D Buhrfeind, and James~W Pennebaker. 1994.
\newblock Expressive writing and coping with job loss.
\newblock \emph{Academy of management journal}, 37(3):722--733.

\bibitem[{Strapparava et~al.(2004)Strapparava, Valitutti
  et~al.}]{strapparava2004wordnet}
Carlo Strapparava, Alessandro Valitutti, et~al. 2004.
\newblock Wordnet affect: An affective extension of wordnet.
\newblock In \emph{Lrec}, volume~4, page~40. Citeseer.

\bibitem[{Vine et~al.(2020)Vine, Boyd, and Pennebaker}]{Vine20Feelings}
V.~Vine, R.L. Boyd, and J.W. Pennebaker. 2020.
\newblock Feelings in many words: Natural emotion vocabularies as windows on
  distress and well-being.
\newblock \emph{Nature Communications}.

\bibitem[{Wiebe et~al.(2005)Wiebe, Wilson, and Cardie}]{wiebe2005annotating}
Janyce Wiebe, Theresa Wilson, and Claire Cardie. 2005.
\newblock Annotating expressions of opinions and emotions in language.
\newblock \emph{Language resources and evaluation}, 39(2-3):165--210.

\end{thebibliography}
\bibliographystyle{acl_natbib}

\clearpage
\appendix
\section*{Appendices}
\label{sec:appendix}

\begin{figure}[h]
    \centering
    \includegraphics[width=\linewidth]{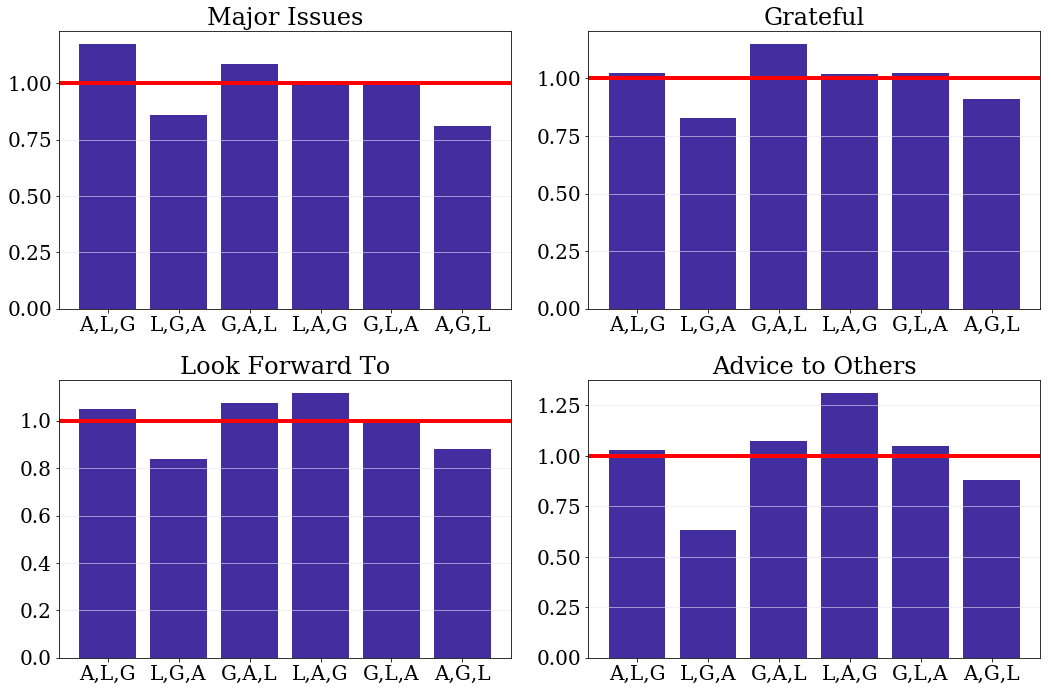}
    \caption{Average number of words in each response grouped by prompt order, divided by the average number of words in each response overall. Equal number of words is at 1, marked with the line. Order of the prompts are indicated by first letter: A = Advice to Others, G = Grateful, L = Looking Forward.}
    \label{fig:length-order-histogram}
\end{figure}

\begin{table*}[h]
    \centering
    \footnotesize
    \begin{tabular}{lrrrrrrr}
    Order & Sessions & Life Sat. & Stress$_{b}$ & Stress$_{a}$  & Personal & Meaning & $\Delta$ Stress \\
    \toprule
Advice to Others,Looking Forward,Grateful & 32 & 5.00 & 3.66 & 3.53 & 5.53 & 5.06 & -0.12 \\
Looking Forward,Grateful,Advice to Others & 29 & 5.10 & 3.62 & 3.17 & 5.31 & 5.21 & -0.45 \\
Grateful,Advice to Others,Looking Forward & 36 & 5.75 & 3.03 & 3.03 & 5.39 & 5.53 & 0.00 \\
Looking Forward,Advice to Others,Grateful & 29 & 5.41 & 3.93 & 2.83 & 5.17 & 5.24 & -1.10 \\
Grateful,Looking Forward,Advice to Others & 22 & 5.27 & 3.73 & 3.59 & 5.05 & 4.77 & -0.14 \\
Advice to Others,Grateful,Looking Forward & 25 & 5.04 & 3.76 & 3.40 & 4.56 & 4.92 & -0.36 \\
    \bottomrule
    \end{tabular}
    \caption{Average ratings grouped by order that the prompts appeared. All sessions begin with ``\textit{Major Issues}."}
    \label{tab:my_label}
\end{table*}

\begin{figure}[h]
    \centering
    \includegraphics[width=\linewidth]{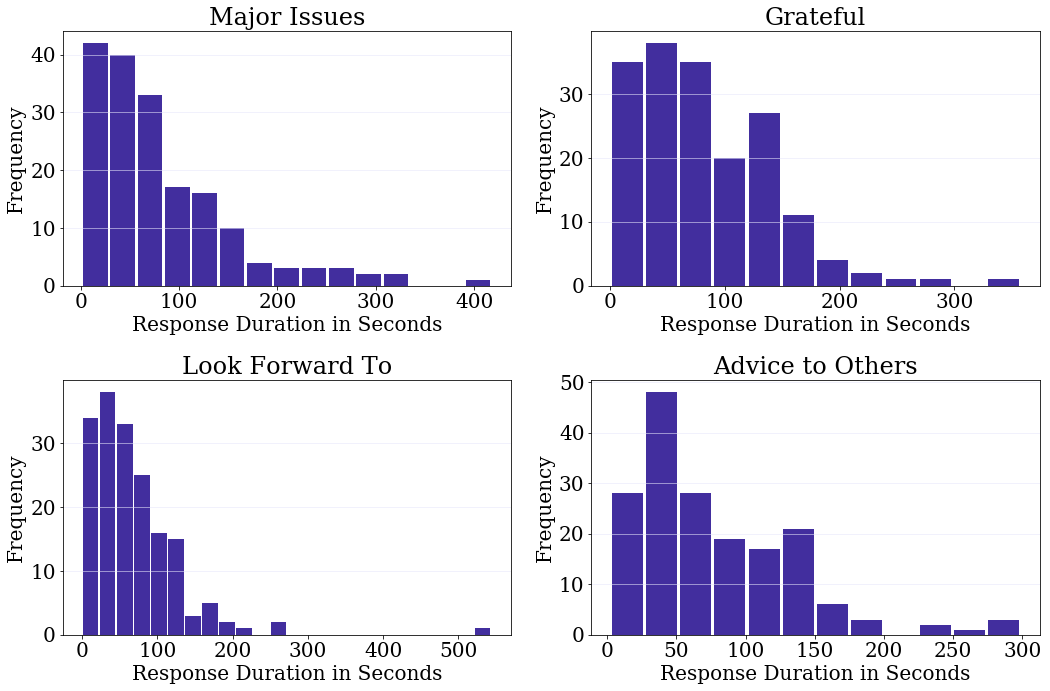}
    \caption{Histogram of the prompt response durations in seconds.}
    \label{fig:duration-histogram}
\end{figure}

\begin{figure}[h]
    \centering
    \includegraphics[width=\linewidth]{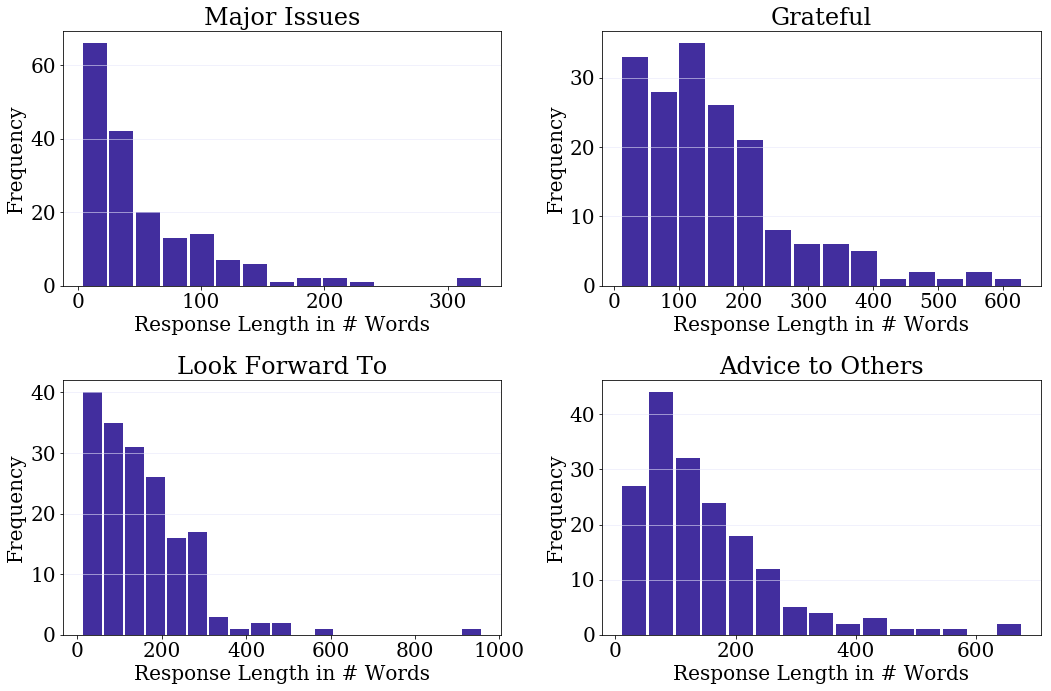}
    \caption{Histogram of the prompt response lengths in tokens.}
    \label{fig:length-histogram}
\end{figure}

\clearpage

\begin{figure}[h]
    \centering
    \includegraphics[width=\linewidth]{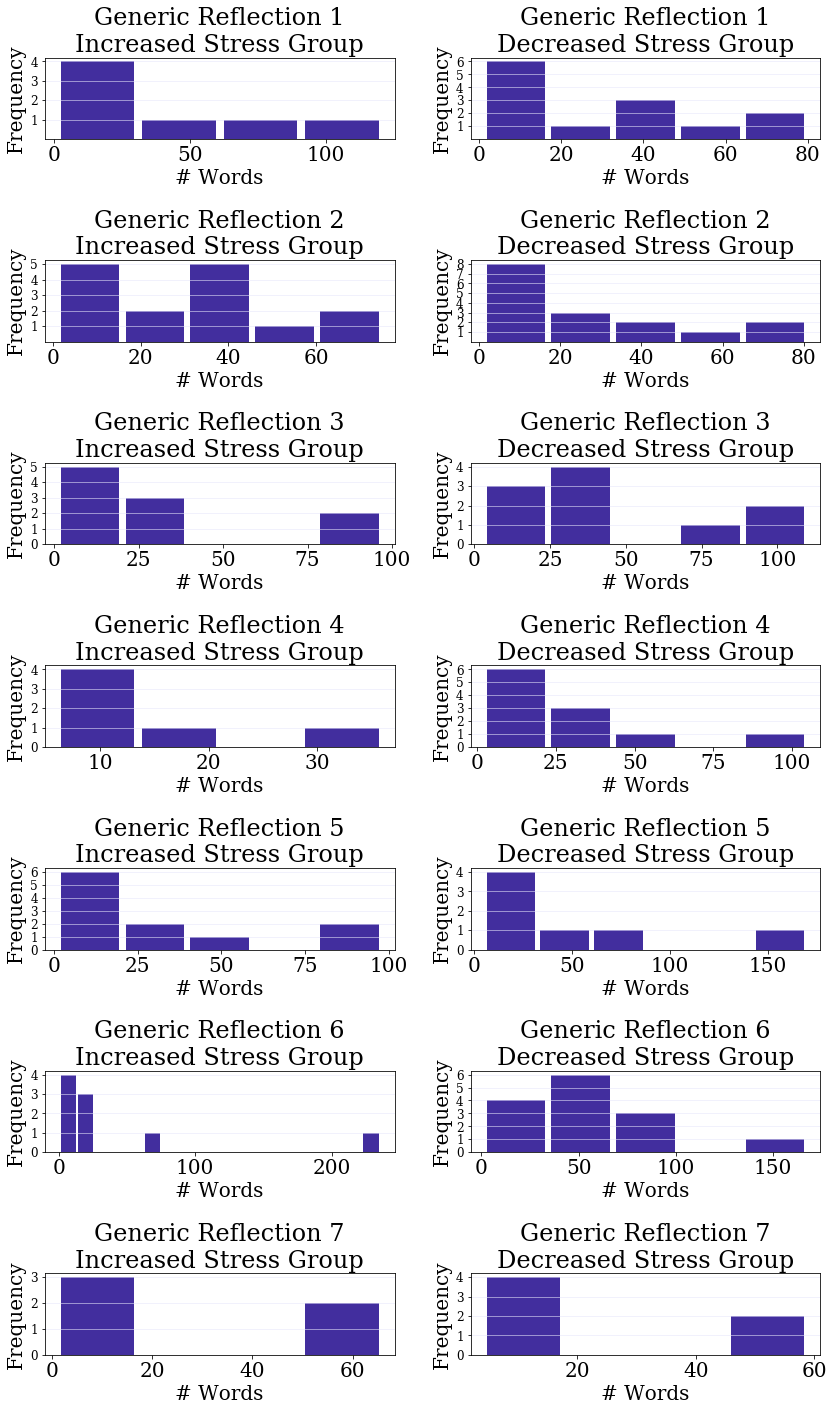}
    \caption{Histograms of the number of words of each user message preceding the generic reflections, grouping users whose stressed increased and decreased.}
    \label{fig:generic-lengths}
\end{figure}
\begin{figure}[h]
    \centering
    \includegraphics[width=\linewidth]{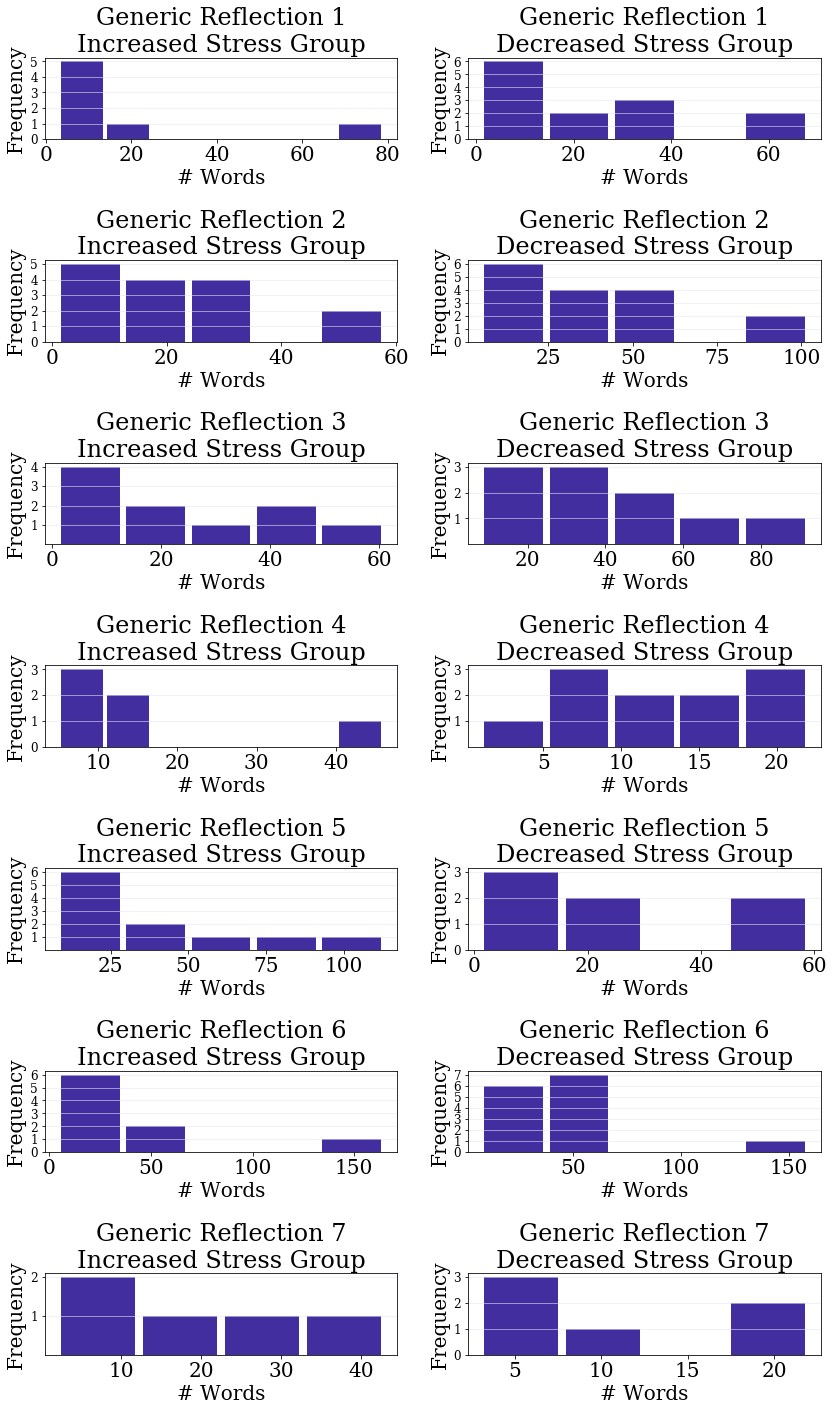}
    \caption{Histograms of the number of words of each user message after the generic reflections, grouping users whose stressed increased and decreased.}
    \label{fig:generic-lengths-after}
\end{figure}

\begin{figure*}[h]
    \centering
    \includegraphics[width=15cm]{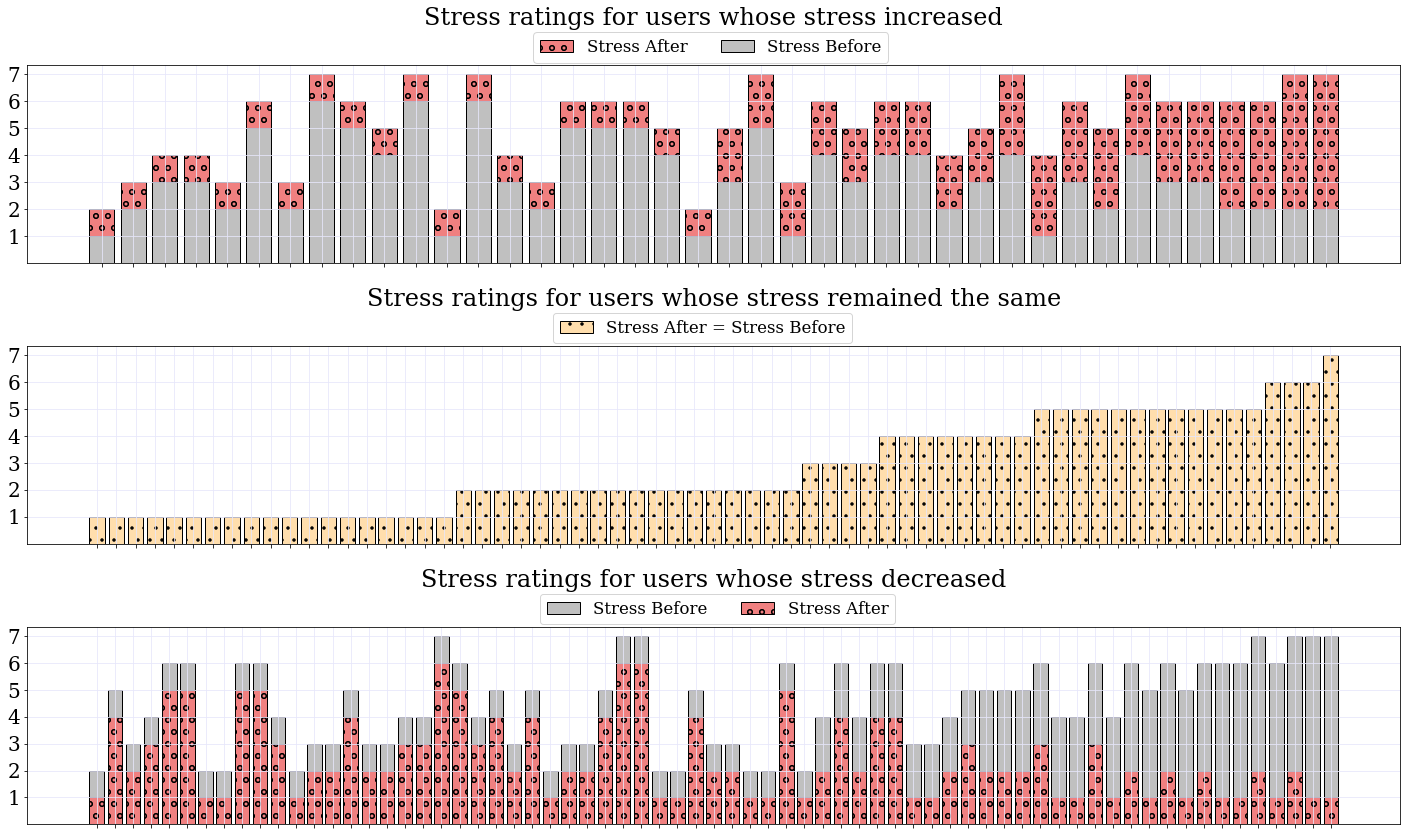}
    \caption{Top: before and after stress ratings by users whose stress increased after interaction with C.P. Middle: before and after stress ratings by users whose stress remained the same after interaction with C.P. Bottom: before and after stress ratings by users whose stress decreased after interaction with C.P. The bars are ordered by the magnitude of change (top and bottom), or by the static stress rating (middle).}
    \label{fig:stress-changes}
\end{figure*}


\end{document}